\documentclass[final,5p,twocolumn]{elsarticle}

\usepackage{amssymb}

\biboptions{sort&compress}

\begin{document}

\begin{frontmatter}

\title{Correlations from hydrodynamic flow in p-Pb collisions}

\author[agh,ifj]{Piotr Bo\.zek}
\ead{Piotr.Bozek@ifj.edu.pl}

\author[ujk,ifj]{Wojciech Broniowski}
\ead{Wojciech.Broniowski@ifj.edu.pl}

\address[agh]{AGH University of Science and Technology, Faculty of Physics and Applied Computer Science, al. Mickiewicza 30, 30-059 Krakow, Poland}
\address[ifj]{The H. Niewodnicza\'nski Institute of Nuclear Physics PAN, 31-342 Krak\'ow, Poland}
\address[ujk]{Institute of Physics, Jan Kochanowski University, 25-406 Kielce, Poland}

\begin{abstract}
Two-particle correlations in relative rapidity and azimuth are studied for the 
p-Pb collisions at the LHC energy of $\sqrt{s_{NN}}=5.02$~TeV in the framework of event-by-event
$3+1$-dimensional  viscous hydrodynamics. It is found that for the highest-multiplicity events the observed ridge 
structures appear in a natural way, suggesting that collective flow may be an important element 
in the evolution of the system. We also discuss the role of the charge balancing and the transverse-momentum
conservation.
\end{abstract}


\end{frontmatter}

Recently, the CMS collaboration has measured the two-dimensional (2D) two-particle correlations in relative pseudorapidity 
and azimuth for the 
proton-Pb (pPb) 
collisions at the energy of $\sqrt{s_{NN}}=5.02$~TeV~\cite{CMS:2012qk}, observing long-range structures, in particular, the
near-side ridge. This fact contributes significantly to the on-going discussion on the very nature of the dynamics and its potential 
collectivity for high-multiplicity systems created in relativistic proton-nucleus  and proton-proton (pp) collisions. 
We recall that in relativistic nucleus-nucleus collisions the ridge structures are naturally explained~\cite{Takahashi:2009na,Luzum:2010sp,Bozek:2012en} 
by the presence of the collective flow arising from hydrodynamics. In~\cite{Bozek:2011if} one of us (PB) has studied the  
hydrodynamic evolution in the most central high-energy pPb and deuteron-Pb collisions, 
with the conclusion that a sizable elliptic and triangular flow can be formed. 

Some degree of collectivity has even 
been suggested for pp collisions~\cite{Luzum:2009sb,d'Enterria:2010hd,Bozek:2009dt,CasalderreySolana:2009uk,Avsar:2010rf,Bozek:2010pb,Werner:2010ss}. 
In high multiplicity pp events a same-side ridge is observed in the 2D correlations functions~\cite{Khachatryan:2010gv}.
This feature could signal the presence of azimuthal correlations in the gluon emission in the initial 
state~\cite{Levin:2011fb,Kovner:2011pe,Kovner:2010xk,Bartels:2011qi,Dusling:2012iga}. The same-side 
ridge observed in pp collisions could also result from the 
azimuthal asymmetry in the collective expansion of the 
small droplet of matter created in the reaction~\cite{Bozek:2010pb,Werner:2010ss}.
 
The hydrodynamic picture in the most central pPb collisions at the LHC is better justified that in pp interactions. The size of the system  is comparable
to the case of most peripheral Pb-Pb collisions, where the elliptic flow has been observed. The initial  density profile and its azimuthal asymmetry in pPb collisions 
can be estimated with the models tested in nucleus-nucleus reactions, whereas the shape of the small fireball that could 
be formed in pp collisions is  less under theoretical control.
In this Letter we model the 2D correlations for the most central pPb collisions in the CMS setup, reproducing the basic features 
of the data, in particular the presence of the near-side ridge. The hydrodynamic description of the fireball, resulting in 
collective flow, may therefore be an efficient approach
even for small colliding systems, bringing in, among other possible sources, a sizable component to the 2D correlations.

The basic object of our study is the two-particle correlation function in relative pseudorapidity and azimuth, defined as~\cite{CMS:2012qk}
\begin{equation}
\label{eq:2pc} \hspace{-4mm}
C_{\rm trig}(\Delta\eta,\Delta\phi) \equiv \frac{1}{N}\frac{d^{2} N^{\rm{pair}}}{d\Delta\eta\, d\Delta\phi}
= B(0,0) \frac{S(\Delta\eta,\Delta\phi)}{B(\Delta\eta,\Delta\phi)},
\end{equation}
where $\Delta\eta$ and $\Delta\phi$ are the relative pseudorapidity and azimuth
of the particles in the pair. The signal is defined with the pairs from the same event,
\begin{equation}
\label{eq:s}
S(\Delta\eta,\Delta\phi) = \langle \frac{1}{N}\frac{d^{2}N^{\rm{same}}}{d\Delta\eta\, d\Delta\phi} \rangle_{\rm events} ,
\end{equation}
while the mixed-event background distribution is
\begin{equation}
\label{eq:b}
B(\Delta\eta,\Delta\phi) = \langle \frac{1}{N}\frac{d^{2}N^{\rm{mix}}}{d\Delta\eta\, d\Delta\phi} \rangle_{\rm mixed \ events}.
\end{equation}
The number of particles $N$ (denoted by CMS as $N_{\rm{trig}}$), is the number of charged particles in 
a given centrality class and acceptance bin, corrected for the experimental efficiency. The introduction of the central bin content, $B(0,0)$, 
brings in the interpretation 
of Eq.~(\ref{eq:2pc}) as the average number of correlated pairs per trigger particle. In our simulations we directly compute the right-hand side of 
Eq.~(\ref{eq:2pc}). 

In pPb collisions, as the small interaction region 
fluctuates widely from event to event, one has to run the costly event-by-event simulations (e-by-e) 
\cite{Takahashi:2009na,Schenke:2010rr,Bozek:2011if,Chaudhuri:2011pa,Petersen:2010cw,Holopainen:2010gz,Qiu:2011hf,Gardim:2011xv,Pang:2012he,Hirano:2012kj}.
It is well known that the inclusion 
of the e-by-e  fluctuations is important for the proper description 
of the initial eccentricity and triangularly, translating into the elliptic and triangular 
flow~\cite{Alver:2008zz,Andrade:2006yh,Alver:2010gr,Alver:2010dn}. For the small pPb system viscosity plays an important role~\cite{Song:2008si}, moreover, the densities are strongly rapidity dependent, hence  viscous $3+1$-dimensional~\cite{Schenke:2010rr,Bozek:2011ua} hydrodynamics 
must necessarily be used. 

\begin{figure}
\includegraphics[angle=0,width=0.485 \textwidth]{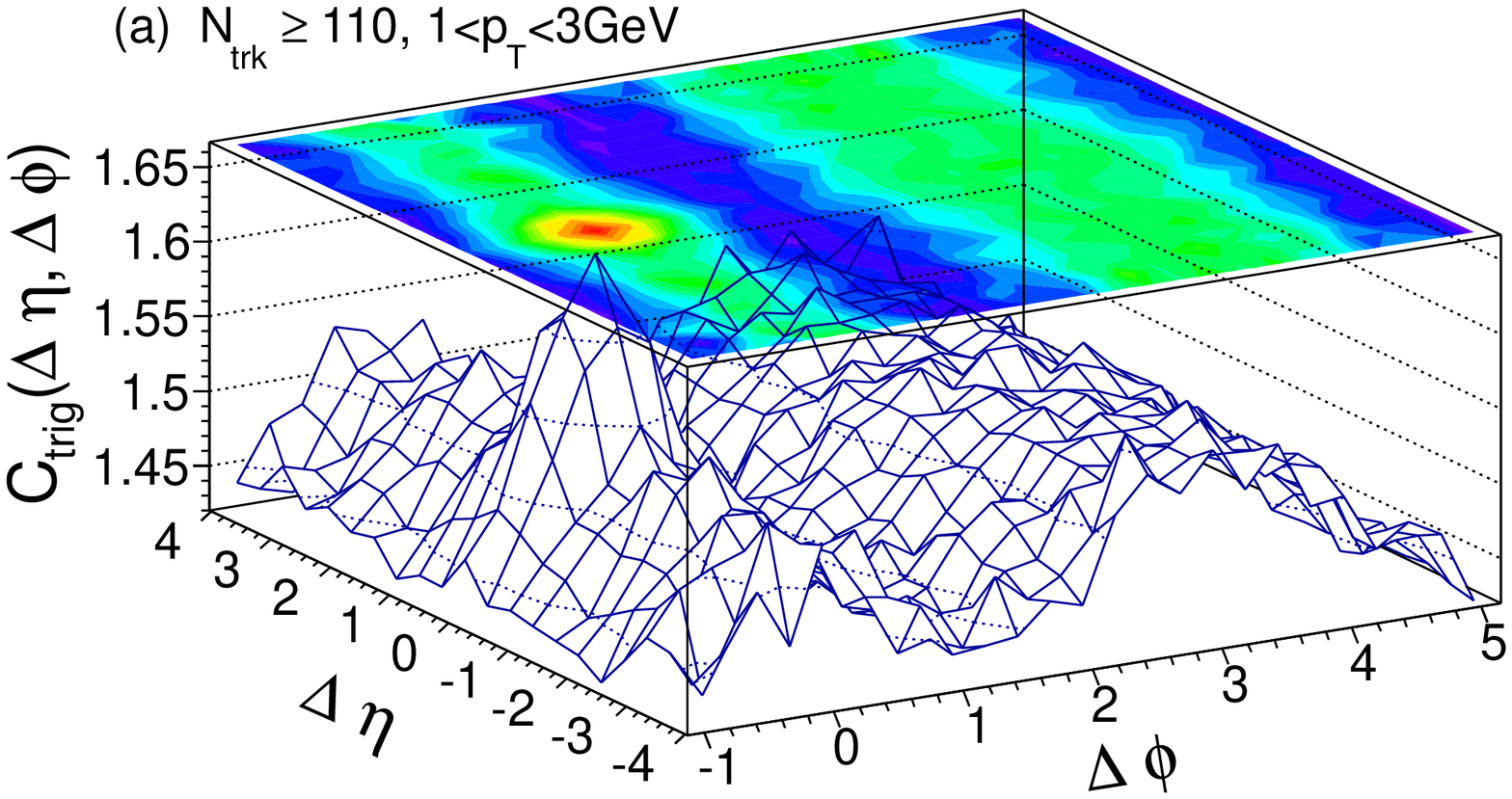} \vspace{-9mm} \\
\includegraphics[angle=0,width=0.485 \textwidth]{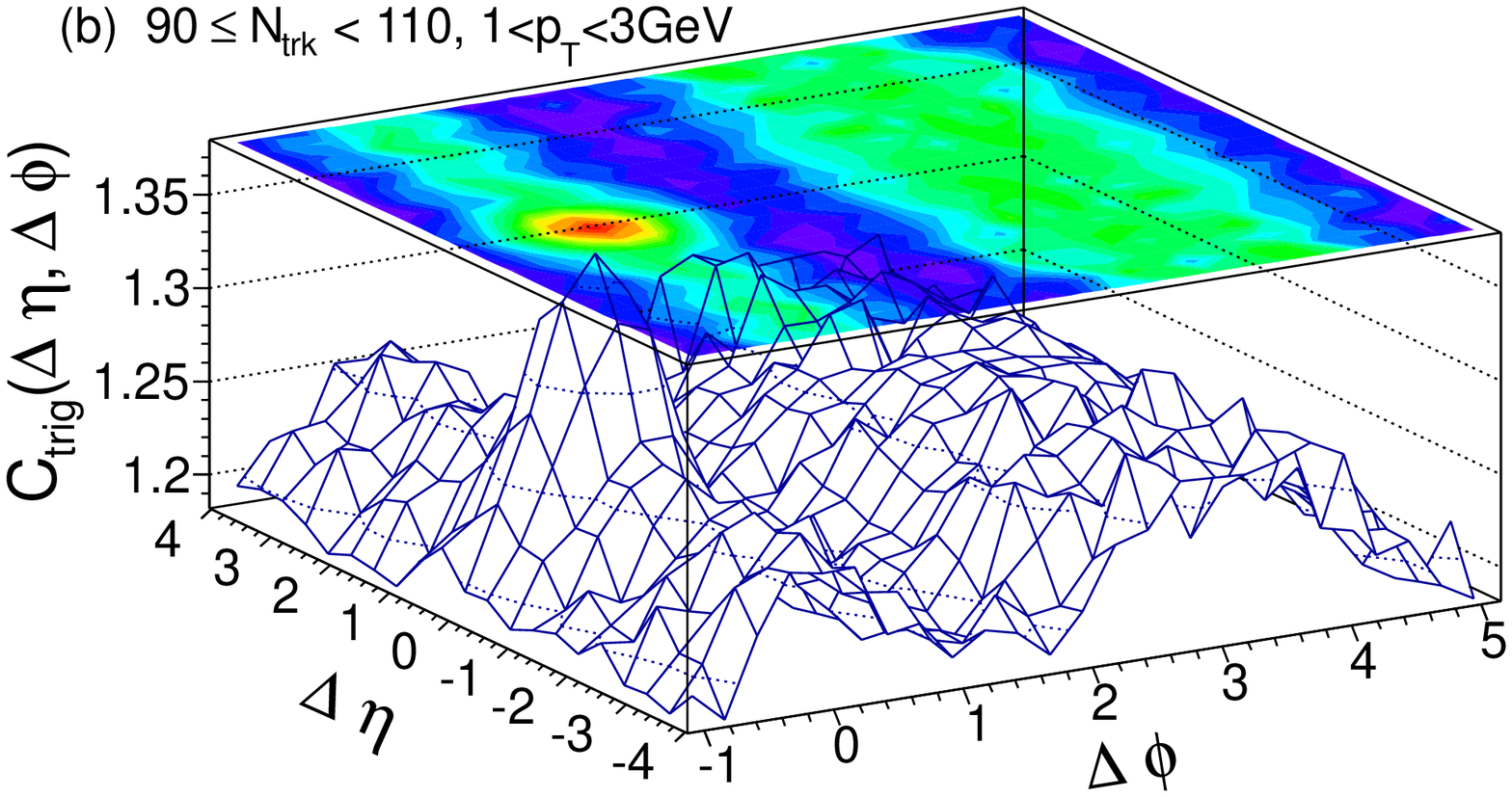} 
\vspace{-7mm}
\caption{The per-trigger-particle 
correlation function $C_{\rm trig}(\Delta\eta, \Delta\phi)$ of Eq.~({\ref{eq:2pc}}) for two most central centrality classes 
of the CMS Collaboration. The transverse momentum of each particle of the pair satisfies $1.0 < p_{T} < 3.0$~GeV.
\label{fig:C}} 
\end{figure}  

We use the  hydrodynamic model  as described in~\cite{Bozek:2011if} to model the 
pPb system at highest centralities. 
The initial condition is generated with GLISSANDO~\cite{Broniowski:2007nz}, implementing various variants 
of the Glauber model~\cite{Glauber:1959aa,Czyz:1969jg,Bialas:1976ed,Bialas:2008zza}.
The parameters of the calculations are similar as in~\cite{Bozek:2011if}, except that they are adjusted for 
the collisions energy of $\sqrt{s_{NN}}=5.02$~TeV.
The nucleon-nucleon cross section is $67.7$~mb, moreover, we use a realistic 
(Gaussian) wounding profile~\cite{Rybczynski:2011wv} for the NN collisions. 
To provide suitable initial conditions for the hydrodynamic evolution, 
the positions of the participants are smoothed with a Gaussian 
with the width of 0.4~fm. 

We use the following initial profile in the space-time rapidity $\eta_\parallel$
\begin{equation}
f(\eta_\parallel)=\exp\left(-\frac{(|\eta_\parallel|-\eta_0)^2}
{2\sigma_\eta^2}\theta(|\eta_\parallel|-\eta_0) \right) \ ,
\label{eq:etaprofile}
\end{equation}
with $\eta_0=2.5$ and $\sigma_\eta=1.4$. The starting time of hydrodynamics is $\tau=0.6$~fm, and the ratio of the shear viscosity to entropy density 
is $\eta/s=0.08$. The expected multiplicity in central pPb collisions is extrapolated linearly in the number of participant nucleons from
the minimum bias results of the ALICE collaboration~\cite{ALICE:2012xs}. Accordingly, the initial entropy per participant in the fireball is adjusted.
For every centrality we produce $300$ initial configurations that are evolved with hydrodynamics to obtain freeze-out hypersurfaces of 
constant temperature $T_f=150$~MeV. Then, for each freeze-out configuration we generate 1000 
THERMINATOR events to efficiently improve the statistics.

\begin{figure}
\begin{center}
\includegraphics[angle=0,width=0.485 \textwidth]{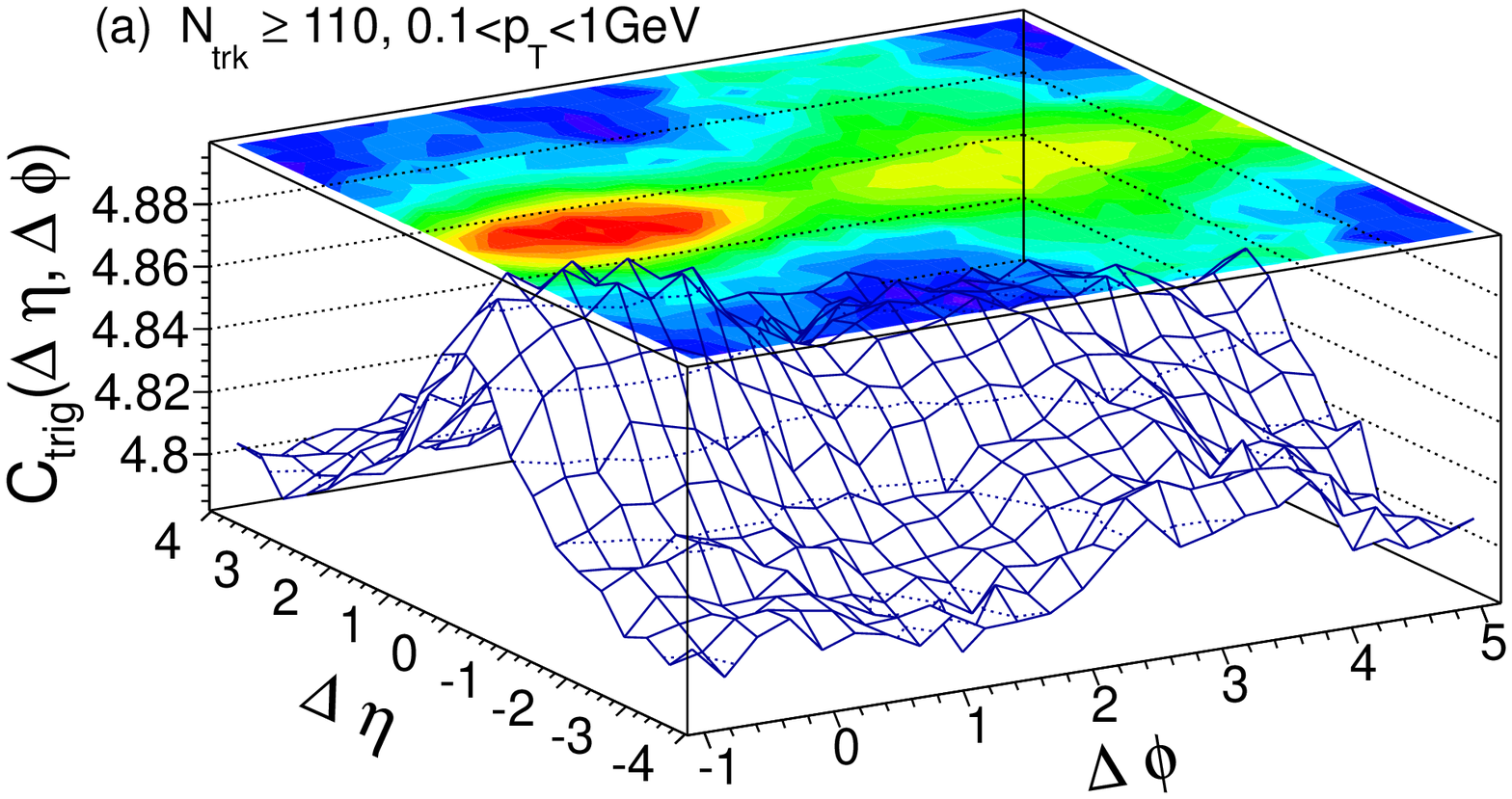} \\ \vspace{-9mm}
\includegraphics[angle=0,width=0.485 \textwidth]{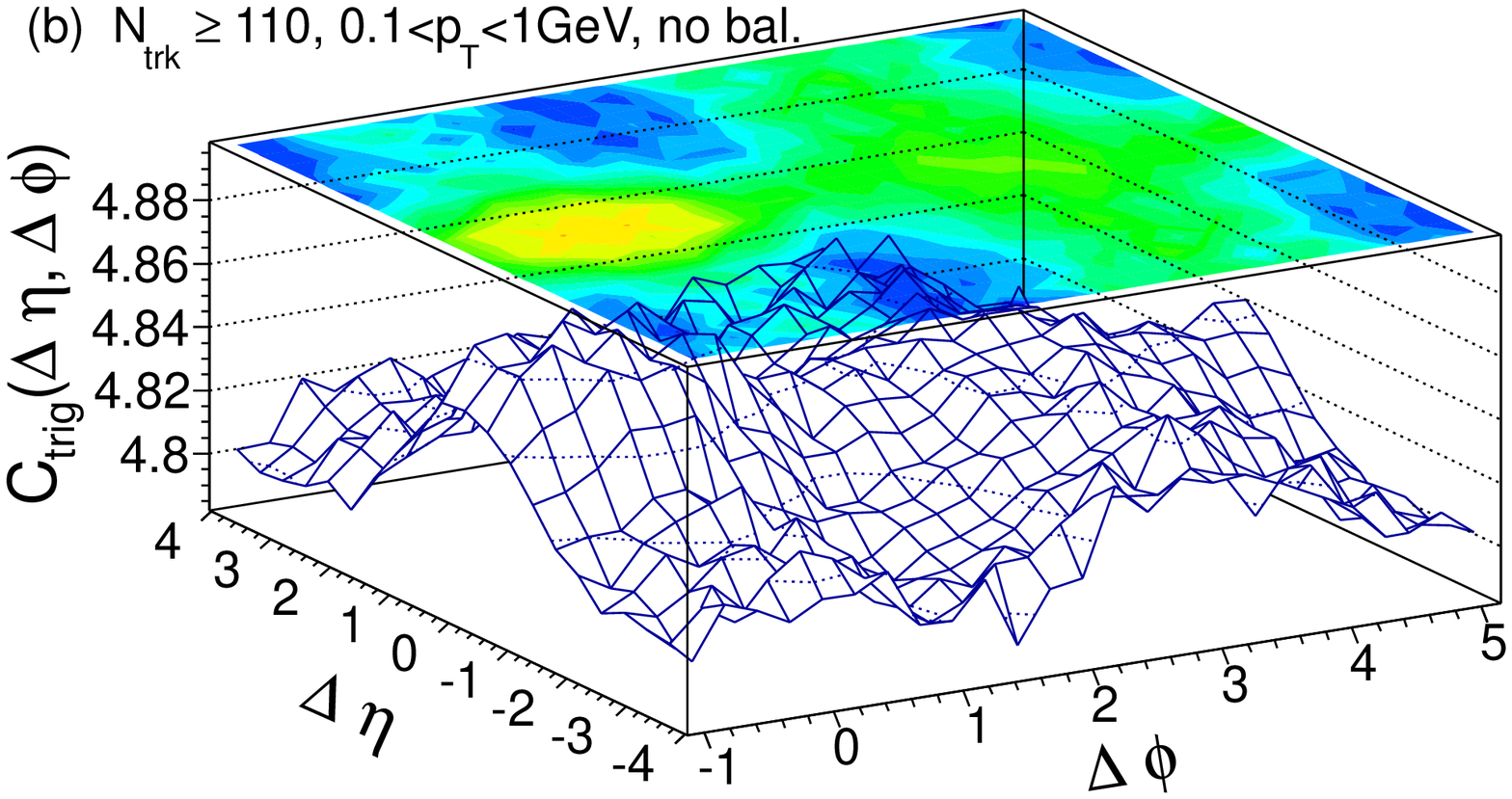} 
\end{center}
\vspace{-7mm}
\caption{The per-trigger-particle correlation function $C_{\rm  trig}(\Delta\eta, \Delta\phi)$ of Eq.~({\ref{eq:2pc}}) for most central events,
$0.1 < p_{T} < 1.0$~GeV. The results including local charge conservation effects at the end of the hydrodynamic evolution are shown 
in panel~(a), whereas panel~(b) presents the case with non-flow correlations from the resonance decays only.
\label{fig:Clow}} 
\end{figure}  

In the statistical emission model the non-flow correlations from resonance decays are built in. Additional 
correlations can appear due to local charge conservation~\cite{Bass:2000az}. Observation indicate that this {\em charge balancing} 
happens at hadronization~\cite{Jeon:2001ue,Bozek:2003qi,Aggarwal:2010ya}, i.e. at the late stage of the evolution. 
The mechanism creates a significant contribution to the 2D
correlations functions~\cite{Bozek:2012en} and is included in the simulations presented below.
Pairs of opposite-charge particles and their antiparticles are emitted locally at 
freeze-out, with a thermal spread in their relative momenta.
Another important source of correlations comes from the global transverse-momentum conservation~\cite{Borghini:2000cm,Bzdak:2010fd}. 
We impose approximately this constraint by requiring that the sum over the particles in  the 
generated event fulfills the condition 
\begin{equation}
\sqrt{\left( \sum_{i} p_x \right)^2 + \left( \sum_i p_y \right)^2}< P_T .
\end{equation}
We have found numerically that limiting the total transverse momentum to $P_T = 5$~GeV is sufficient; 
further reduction does not affect the studied quantities. This amounts for 
retaining about 8\% of the least-$P_T$ events from our sample. 

We apply the hydrodynamic model to the two most central centrality classes used by the CMS Collaboration.
The centrality of the events is defined based on the charged particle multiplicity in the CMS acceptance.
A good approximation of the centrality cuts in our model is represented by simple conditions on the number of 
the participant nucleons. The most central collisions with $N_{part}\ge 18$ amount to  $3.4\%$ of most central events
in the Glauber Monte Carlo model. The second most central class is defined by $16 \le N_{part} \le 17$ and
sums up $4.4\%$ of the cross section. Cuts on the final multiplicity in the calculations instead of $N_{part}$ 
could be used, once the model of the initial state were supplemented with effects of fluctuations of the 
energy deposited in each elementary collision~\cite{Dumitru:2012yr,Tribedy:2010ab,Moreland:2012qw,Schenke:2012hg}, 
but this is not crucial for our study.

In the hydrodynamic model the multiplicity fluctuations are largely decoupled from the collective expansion phase.
Our model  gives realistic predictions on the collective flow, but the multiplicity distribution cannot
be reliably calculated. This has a consequence for the normalization of the correlation functions.
By integrating the per trigger correlation function one obtains
\begin{equation}
\int d\Delta \phi d\Delta \eta \; C_{\rm  trig}(\Delta\eta,\Delta\phi)=\frac{\langle N(N-1)\rangle}{\langle N \rangle} \ ,
\end{equation}
i.e., the ratio of the average number of pairs over average multiplicity in a given acceptance window. 
In the presence of correlations from 
collective flow only, a more robust observable is the 2D correlation function normalized by the number of 
pairs instead of $N$ in Eq.~(\ref{eq:2pc}). 

In the hydrodynamic model collective flow dominates in the 
correlation function for $2<|\Delta \eta|<4$. Therefore to make a meaningful estimate of the 
hydrodynamic component in the 2D correlation function, we rescale the calculated functions to get the same
subtraction constant $C_{\rm ZYAM}$ in the zero-yield-at-minimum (ZYAM) procedure. We use the ZYAM values as quoted by the CMS Collaboration
for each multiplicity and $p_T$ bin~\cite{CMS:2012qk}. Such rescaled correlation functions, 
called normalized correlation functions in the following, 
should be used to estimate the contribution of the collective flow 
to the ridge observed in the experiment.

In the following we describe the results obtained with our simulations. We begin with the 
correlation function of Eq.~({\ref{eq:2pc}}), shown in Fig.~\ref{fig:C} for the most 
central collisions with two different cuts imposed on the transverse momentum of each particle 
in the pair. 
The 2D correlations function presents similar features as the experimental one~\cite{CMS:2012qk}. A sharp same-side peak
is formed due to the resonance decays and the local charge conservation~\cite{Bozek:2012en}. The observed  additional 
correlations from jet fragmentation at small $\Delta \phi$-$\Delta \eta$, or the Bose-Einstein and Coulomb correlations, 
are not included in our model.
The away- and same-side ridges are formed in the whole range of $\Delta \eta$. The shape of these ridges is determined
mainly by the first 3 harmonics in the relative azimuth. The first harmonic comes predominantly from the 
transverse-momentum conservation and is seen as a tendency for the back-to-back emission. The second and third harmonics 
are provided by the collective expansion of the initial fluctuating source and describe well the shape and the width of the
same- and away-side ridges. As expected~\cite{Bozek:2010pb,Werner:2010ss}, the collective elliptic flow
leads to the formation of the same-side ridge in the 2D correlation functions, which is our basic observation.

\begin{figure}
\hspace{-11mm} \includegraphics[angle=0,width=0.55 \textwidth]{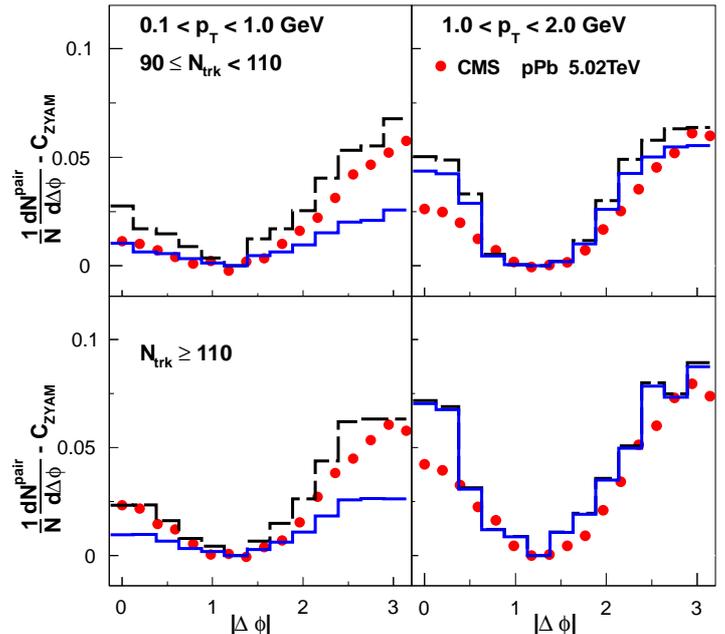} 
\vspace{-7mm}
\caption{The projected and ZYAM-subtracted correlation function in the region $2<|\Delta \eta|<4$ for the two most  
central bins in multiplicity (panels extending horizontally)
and two $p_T$ intervals (panels extending vertically) for the pPb collisions. 
The CMS measurement~\cite{CMS:2012qk} is shown as dots.  The results 
of our hydrodynamic model with the normalized correlation functions are shown with the solid lines. 
The dashed lines show the results of the hydrodynamic model with subtraction of the model ZYAM values and no rescaling. 
\label{fig:P}} 
\end{figure}  

A qualitatively different behavior is visualized in Fig.~\ref{fig:Clow}. At low $p_T$ 
the correlation displays a ridge (panel~a) in the azimuthal angle direction
(near $\Delta\eta=0$), which is due to
charge balancing in the hadronization and resonance decays. 
Since the harmonic flow is small at low $p_T$, only weak traces of the ridges (near $\Delta \phi=0$ and 
$\Delta \phi=\pi$) are visible. To evidence the effect of the local charge conservation in the formation of 
the $\Delta \eta \simeq 0$ ridge, we show in panel~b the correlation obtained without charge 
balancing. The structure is much less pronounced now, as it is due to resonance decays only.
We note, interestingly, that similar structures, with a ridge in the $\Delta \phi$ direction in the 2D correlation functions, have been observed 
in pp collisions at $7$~TeV by the ALICE Collaboration~\cite{mjanikwpcf}.

In the higher~$p_T$ bins (Fig.~\ref{fig:C}), the average transverse flow forms prominent ridges 
at $\Delta \phi \simeq 0$ and $\Delta \phi \simeq \pi$ which hide the  $\Delta \eta \simeq 0$ ridge.
Also, as the flow increases, the unlike-sign pairs become collimated in the azimuthal angle~\cite{Bozek:2004dt,Bozek:2012en} which 
contributes to the rise of the central peak. 

In a small system, a substantial part 
of particles is expected to be emitted from the corona, without 
subsequent rescattering~\cite{Bozek:2005eu,Hohne:2006ks,Werner:2007bf,Becattini:2008ya}. Particles emitted from the corona give
a separate contribution to the 2D correlation function. In particular, one expects less collimation from
charge balancing~\cite{Bass:2000az}. The shape of the $\Delta \eta \simeq 0 $ ridge could be used
to separate the contributions from the thermalized core and the corona in  pPb reactions.
In the kinematic region $|\Delta\eta|>2$, selected for the analysis of 
the projected correlation function $C_{\rm  trig}(\Delta \eta, \Delta \phi)$,
the short-range charge balancing effects are not important. 

Next, in order to compare quantitatively to the CMS data~\cite{CMS:2012qk}, we show in Fig.~\ref{fig:P} the averaged correlation 
function  
\begin{eqnarray}
\frac{1}{N} \frac{dN^{\rm pair}}{d\Delta \phi} = \int_{2 <|\Delta\eta|<4} \!\!\!\!\!\!\!\!\! d\Delta \eta \; C_{\rm  trig}(\Delta \eta, \Delta \phi)/\int_{2 <|\Delta\eta|<4} \!\!\!\!\!\!\!\!\! d\Delta \eta. \label{eq:P} 
\end{eqnarray}
We note that our e-by-e hydrodynamic simulations (lines) have the desired two-ridge 
structure, which is generated by the harmonic flow. 
The incorporated transverse momentum conservation increases the relative strength of the away-side 
ridge. The normalization of the correlation function 
is chosen to reproduce the normalization in the experiment. The function 
$C_{\rm  trig}(\Delta \eta, \Delta \phi)$ is thus rescaled to obtain the same value of the parameters $C_{\rm ZYAM}$ as for
the data points in Fig.~2 of~\cite{CMS:2012qk}. The normalization procedure assures that the ratio
$\langle  N(N-1) \rangle>/<\langle N>$  is the
similar as in the experiment. These normalized results, denoted with the 
solid lines in Fig.~\ref{fig:P} are in good agreement for the $1.0 < p_T < 2.0 $~GeV case, and reproduce 
part of the ridge amplitudes in the lowest $p_T $ bin, $0.1 < p_T < 1.0 $~GeV. The deviations may appear for 
many different reasons, namely, other sources of correlations are present, the initial eccentricities calculated 
in the Glauber model without
fluctuation at sub-nuclear scales may be underestimated, or contributions from the non-thermal corona in the interaction region are important.
In Fig. \ref{fig:P} are also shown the results obtained without normalizing the correlations functions 
but just subtracting the values at the minimum of the yield (dashed lines), $C_{\rm ZYAM}$; in that case the parameters
$C_{\rm ZYAM}$ are a factor $1.1$ to $2.5$ larger than in the experiment. Thus the better agreement with 
the experimental points of the dashed lines is partly accidental, due to a mismatch in the particle multiplicities. 
Nevertheless, we show these curves, as the issues related to proper 
normalization between experiment and a model are far from trivial.

\begin{figure}
\begin{center}
\includegraphics[angle=0,width=0.41 \textwidth]{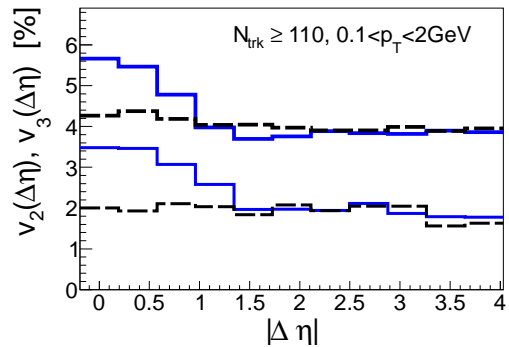} 
\end{center}
\vspace{-4mm}
\caption{The flow coefficients $v_2(\Delta \eta)$ (upper lines) and $v_3(\Delta \eta)$ (lower lines) calculated
from the two-particle correlations as function of the relative pseudorapidity of the particles in the pair.
The solid and dashed lines are for the unlike- and like-sign pairs, respectively. The central peak is due to 
charge balancing and, to a lesser extent, resonance decays.
\label{fig:v23}} 
\end{figure}  

In view of the recent results of the CMS Collaboration for the 2D correlations functions in pPb collisions, it is
interesting to look at the possibility of measuring directly the harmonic flow coefficients. We plot the elliptic
and triangular flow coefficients as function of the pseudorapidity 
gap in Fig.~\ref{fig:v23}. The quantities are obtained in our hydrodynamic model 
from the Fourier decomposition of the correlation function $C_{\rm trig}(\Delta \eta, \Delta \phi)$~\cite{Bozek:2012en}. 
The non-flow effects present in our model are important only for pairs 
of small pseudorapidity separation. In the intervals $2<|\Delta \eta|<4$ the non-flow effects
from the resonance decays and the local charge conservation can be neglected. We note that the 
flow coefficients in Fig.~\ref{fig:v23} are sizable, thus
could be measured. It must be noted, however, that other sources of non-flow correlations  may be present
also in that kinematic region, but with smaller amplitudes, as measured in pp collisions~\cite{Khachatryan:2010gv}.

In this Letter we have explore the possibility 
that the azimuthally asymmetric collective flow is generated in the hydrodynamic expansion 
of the fireball created in pPb collisions~\cite{Bozek:2011if}. The flow asymmetry together with 
the transverse-momentum correlations is capable of reproducing the observed form of the 2D correlation 
functions in $\Delta \eta$-$\Delta \phi$ for the studied two highest multiplicity and two lowest $p_T$ bins, where
hydrodynamics is expected to apply. The agreement is semiquantitative, but very suggestive. 
The found important contribution of 
correlations of hydrodynamic origin (flow) does not exclude the presence of other sources of correlations 
between emitted particles, such as jets, string decays, or correlated gluon emission. We also note that 
the hydrodynamic description can be improved in many ways, including fluctuations at sub-nuclear scales, 
the  early flow, contributions from the corona, or varying the viscosity coefficients. 
The significant increase of the ridge amplitude when going from pp to pPb collisions indicates that 
the harmonic coefficients of the collective flow become sizable and could be directly measured in spite of 
the background of the non-flow correlations in the small system.

We notice the formation of the ridge at $\Delta \eta \simeq 0$ at low transverse momenta in our
pPb simulations, similarly to the recent findings of~\cite{mjanikwpcf}. 

\bigskip


Supported by Polish Ministry of Science and Higher Education, grant N~N202~263438, and by 
National Science Centre, grant DEC-2011/01/D/ST2/00772.

\bibliography{../hydr}

\end{document}